\documentclass {article}

\usepackage{amsmath,amssymb}
\usepackage{verbatim}  

\title {Comment on ``The Cosmic Time in Terms of the Redshift'', by Carmeli et al.}
\author
{Alan Macdonald \\
Department of Mathematics \\
Luther College, Decorah, IA 52101,  U.S.A. \\ 
macdonal@luther.edu}
\begin {document}
\maketitle
\begin{abstract}
\noindent
The time-redshift relation of Carmeli et al. differs from that of the standard flat $\Lambda$CDM model by more than 500 million years for $1 \le z \le 4.5$.
\end{abstract}

\vspace{.1in}
In a recent paper in this journal,
Carmeli et al. investigated the relationship between the time $t$ that light from a distant object was emitted and its redshift $z$ \cite{Carmeli}.
They ``start by assuming that the universe is empty of gravitation'' 
and derive the time-redshift relation in this case:
\begin{align} \label{eq:flat}
t(z) = \frac{2t_0}{1 + (1+z)^2}\,,
\end{align}
where $t_0$ is the age of the universe. (Carmeli et al. denote the age by $\tau$.)

Then they modify Eq. (\ref{eq:flat}) to include gravity. 
They ``adopt the method used in classical general relativity theory'' and replace $t_0$ by $H_0^{-1}$:
\begin{align}\label{eq:Carmeli}
t(z) = \frac{2H_0^{-1}}{1 + (1+z)^2}\,,
\end{align}
because it ``seems to be the proper thing to do''. 

Carmeli et al. write that Eq. (\ref{eq:Carmeli}) is ``valid for all redshift values'' and
``It is hoped that the formula will be useful for identifying objects at the early Universe.''
They nowhere restrict their equation to specific general relativistic cosmological models.

The purpose of this comment is to point out that Eq. (\ref{eq:Carmeli}) is not a good approximation to the time-redshift relation for the currently favored standard flat $\Lambda$CDM cosmological model.
Specifically, they differ by over half a billion years for $1 \le z \le 4.5$. 

We can see immediately that Eq. (\ref{eq:Carmeli}) is not an exact time-redshift relation. 
For it does not give $t(0) = t_0$, which is obviously true for every cosmological model. 
(Eq. (\ref{eq:flat}) does pass this test.)

Gr{\o}n \cite{Gron} has given a closed form expression for the expansion factor $S(t)$ in the standard flat $\Lambda$CDM model:
\begin{align}
S\,^3(t) = \frac{\Omega_M}{\Omega_\Lambda}
                      \sinh^2 \left( \frac{3H_0 \Omega_\Lambda^{\,1/2}} {2}\, t \right) S\,^3(t_o),
\end{align}
where $H_0$ is Hubble's constant,
$\Omega_M$ is the fractional matter energy density today,
and $\Omega_\Lambda$ is the fractional dark energy density today.

Combining this with $z + 1 = S(t_o)/S(t)$ gives	the time-redshift relation	
\begin{align}\label{eq:Gron}
t(z) = \frac{2H_0^{-1}}{3\,\Omega_\Lambda^{\,1/2}}
    \sinh^{-1}\left[ \left(\frac{\Omega_\Lambda}{\Omega_M}\right)^{\!1/2} (z + 1)^{-3/2} \right].
\end{align}

The currently accepted best values of the parameters in Eq. (\ref{eq:Gron}) are 
$H_0 = 72$ km/sec/Mpc, $\Omega_M = .26$, and $\Omega_\Lambda = .74$ \cite{Spergel}.
(There is no consensus about the nature of dark matter and dark energy. 
But there is a wide consensus about the values of $\Omega_M$ and $\Omega_\Lambda$.)
Substituting these values shows that the difference between Eqs. (\ref{eq:Carmeli}) and (\ref{eq:Gron}) 
is over half a billion years for $1 \le z \le 4.5$. 

{}


\begin{thebibliography}{}
\small
\bibitem{Carmeli}M. Carmeli, J. Hartnett, F. Oliveira, ``The Cosmic Time in Terms of the Redshift'',
Found. Phys. Lett. {\bf 19} 277-283 (2006). Also gr-qc/0506079.

\bibitem{Gron}\O. Gr{\o}n, ``A new standard model of the universe'', Eur. J. Phys. {\bf 23} 135-144 (2002).

\bibitem{Spergel}D. Spergel, et al., ``Wilkinson Microwave Anisotropy Probe (WMAP) Three Year Results: Implications for Cosmology'', ApJ submitted, astro-ph/0603449.

\end{thebibliography}
\end{document}